\documentclass[aps,preprint,preprintnumbers,amsmath,amssymb,nofootinbib]{revtex4}
\usepackage{amssymb}
\usepackage{lscape}
\usepackage{amsmath}
\usepackage{bm}
\usepackage{graphicx}
\usepackage{epsfig}
\begin{document}

\title{Origin of time before inflation from a topological phase transition}
\author{$^{1,2}$ Mauricio Bellini\footnote{E-mail address: mbellini@mdp.edu.ar} }
\address{$^1$ Departamento de F\'isica, Facultad de Ciencias Exactas y
Naturales, Universidad Nacional de Mar del Plata, Funes 3350, C.P.
7600, Mar del Plata, Argentina.\\
$^2$ Instituto de Investigaciones F\'{\i}sicas de Mar del Plata (IFIMAR), \\
Consejo Nacional de Investigaciones Cient\'ificas y T\'ecnicas
(CONICET), Mar del Plata, Argentina.}

\begin{abstract}
We study the origin of the universe (or pre-inflation) by suggesting that the primordial space-time in the universe suffered a global topological phase transition, from a
4D Euclidean manifold to an asymptotic 4D hyperbolic one. We introduce a complex time, $\tau$, such that its real part becomes dominant after started the topological phase transition. Before the big bang, $\tau$ is a space-like coordinate, so that can be considered as a reversal variable. After the phase transition is converted in a causal variable. The formalism solves in a natural manner the quantum to classical transition of the geometrical relativistic quantum fluctuations: $\sigma$, which has a geometric origin.\\
\end{abstract}
\maketitle

\section{Introduction and Proposal}

The inflationary theory of the universe provides a physical mechanism to
generate primordial energy density fluctuations on cosmological scales \cite{infl}. The primordial
scalar perturbations drive the seeds of large scale structure which then had gradually formed today's galaxies, which is being
tested in current observations of cosmic microwave background (CMB). These fluctuations are today larger than a thousand size of a typical galaxy, but during inflation
were very much larger than the size of the causal horizon. According with this scenario, the almost constant potential depending of a minimally coupled to
gravity inflation field $\varphi$, called the inflaton, caused the accelerated expansion of the very early universe. The cosmic microwave background that we observe today is almost
isotropic. The background temperature is about 2.7 K with a tiny fluctuation at a level of about $10^{-5}$ K. This is consistent
with measurements of matter structures in the universe at cosmological scales, where the universe is almost homogeneous.

The theory that describes the earlier evolution of the universe is called pre-inflation\cite{Kur}.
The existence of a pre-inflationary epoch with fast-roll of the inflaton field would introduce an
infrared depression in the primordial power spectrum. This depression might have left an imprint in the CMB anisotropy\cite{GR}.
It is supposed that during pre-inflation the universe begun to expand from some Planckian-size initial volume to thereafter pass to an inflationary epoch.
Some models consider the possibility of an pre-inflationary epoch in which the universe is dominated by radiation\cite{IC}.
the metric fluctuations can be studied
as a more profound phenomena in which the scalar metric
fluctuations appear as a geometric response to the scalar field
fluctuations by means of geometrical displacement from a Riemann
manifold to a Weylian one, through RQG.  The dynamics of the geometrical scalar field is defined on a Weyl-integrable manifold that preserves the gauge-invariance under the transformations of the Einstein's equations, that involves the cosmological constant. Our approach is different to quantum gravity.
The natural way to construct quantum gravity models is to apply quantum field theory methods to the theories of classical gravitational
fields interacting with matter. Our approach is different to quantum gravity because our subject of study is the dynamics of the geometrical quantum fields\cite{rb}.

With the aim to describe the origin of the universe during pre-inflation we shall consider a complex manifold, in terms of which the universe describes a background semi-Riemannian expansion, with a line element
for this case is
\begin{equation}\label{1}
d\hat{S}^2 = \hat{g}_{\mu\nu} d\hat{x}^{\mu} d\hat{x}^{\nu}= e^{2i\theta(t)} d\hat{t}^2 + a^2(t) \hat{\eta}_{ij} d\hat{x}^i d\hat{x}^j,
\end{equation}
with the signature: $(+,-,-,-)$. Here $\theta(t)=\frac{\pi}{2} \frac{a_0}{a}$, with $a\geq a_0$, $t$ is a real parameter time and $H_0=\pi/(2 a_0)=1/t_p$, such that $t_p=5.4 \times 10^{-44} \, {\rm sec}$ is the Planckian time. Notice that the metric (\ref{1})
describes a complex manifold such that, at $t=0$ the space-time is Euclidean, but after many Planckian times, when $\theta \rightarrow 0$, it becomes hyperbolic.
We shall define the background action ${\cal I}$ on this manifold, so that it describes the expansion driven by a scalar field, which is minimally coupled to gravity
\begin{equation}
{\cal I} = \int \, d^4x\, \sqrt{\hat{g}}\, \left[ \frac{{\cal \hat{R}}}{16 \pi G} +  \left[ \frac{1}{2}\dot\phi^2 - V(\phi)\right]\right],
\end{equation}
where $\sqrt{\hat{g}} = i a^3 e^{i\theta}$. The metric (\ref{1}) is not sufficiently explicit to describe the transition to an inflationary universe from a topological phase transition, because $t$ is not exactly
the dynamical coordinate that describes this transition.

\section{Origin of time in the universe}

\begin{figure}[h]
\noindent
\includegraphics[width=.6\textwidth]{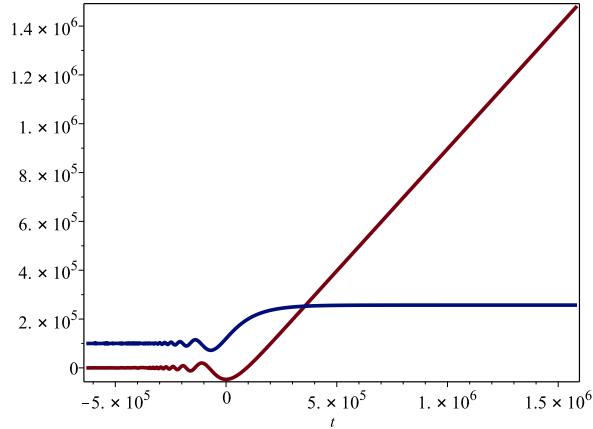}\vskip -3cm\caption{Plot of $\tau(t)$. The red line describes the real part of $\tau$, and the blue line shows the imaginary part of $\tau$ for $H_0=10^{-5}\,{\rm G^{-1/2}}$. Before the big bang, $\tau$ is a space-like coordinate, so that can be considered as a reversal variable. However, after the phase transition it changes its signature and then can be considered as a causal variable in the framework of a 4D space-time.}\label{time}
\end{figure}

The correct dynamical variable in (\ref{1}) is: $\tau=\int e^{i \hat\theta(t)} dt$, which describes the time on the complex plane.
The real part of $\tau$ is the causal time, that increases linearly with $t$ for $t> 1/H_0$. However, it is oscillating around zero, with increasing amplitude, for $t<0$. On the other hand, the imaginary part of $\tau$ oscillates around a fixed value for $t<0$, but it remains constant for $t>1/H_0$. In other words, for $t>1/H_0$ the imaginary part of $\tau$ is freezed, so that its real part increases and dominates to determinate the hyperbolicity of space time. The plot of the real and imaginary parts of $\tau$ can be seen in the Fig. (\ref{time}). Notice that for $t\gg1/H_0$, we obtain that $\tau \rightarrow t$. The idea is that $\tau$ to be a space-like coordinate before the big bang, so that it can be considered as a reversal variable. However, after a topological phase transition we must require that it changes its signature and then can be considered as a causal (irreversible) variable.

We shall consider a scale factor, related to a de Sitter expansion in the $t$-dynamical scale: $H_0=\dot{a}/a(t)$, such that ${\cal H}(\tau)= \frac{1}{a(\tau)} \frac{d a(\tau)}{d\tau}=H_0\, e^{-i\hat{\theta}(\tau)}$, is
\begin{equation}\label{sf}
a(\tau) = a_0\, e^{{\rm Ei}\left[1, i\frac{\pi}{2} a_0 e^{-H_0 \tau}\right]}.
\end{equation}
Notice that we have used the fact that $\hat{\theta}(\tau) = \frac{\pi}{2} \frac{a_0}{a(\tau)}$. The expression (\ref{sf}) for the scale factor written in terms
of $\tau$ makes very difficult to describe the cosmological dynamics of the universe. For this reason, we shall search another variable to describe the dynamics of this topological phase
transition.

\section{New dynamical variable and pre-inflation}

A good candidate is the phase $\hat{\theta}$. Since $\hat\theta(t)=\frac{\pi}{2} e^{-H_0 t}$, we can rewrite the metric (\ref{1}), as
\begin{equation}\label{m}
d\hat{S}^2 =  \left(\frac{\pi a_0}{2}\right)^2 \frac{1}{\hat{\theta}^2} \left[{d\hat{\theta}}^2 + \hat{\eta}_{ij} d\hat{x}^i d\hat{x}^j\right].
\end{equation}
If we desire to describe an initially Euclidean 4D universe, that thereafter evolves to an asymptotic value $\hat{\theta} {\rightarrow } 0$, we must require that $\hat{\theta}$ to be with an initial value $\hat\theta_0=\frac{\pi}{2}$. Furthermore, the nonzero components of the Einstein tensor, are
\begin{equation}
G_{00} = - \frac{3}{\hat{\theta}^2} , \qquad G_{ij} =  \frac{3}{\hat\theta^2 } \,\delta{ij},
\end{equation}
so that the radiation energy density and pressure, are respectively given in this representation by $\rho(\hat\theta) = \frac{1}{2\pi G} \frac{3}{(\pi a_0)^2}$, $P(\hat\theta) = - \frac{1}{4\pi G} \frac{3}{(\pi a_0)^2}$. The equation of state for the metric (\ref{m}), describes an vacuum expansion: $\omega(\hat\theta) =  - 1$. In this case the asymptotic scale factor, Hubble parameter and the potential are are respectively given by
\begin{equation}
a(t)= a_0\, e^{H_0 t}, \qquad \frac{\dot{a}}{a} = H_0 \qquad V= \frac{3}{8\pi G} H^2_0,
\end{equation}
so that the background field solution is given by a constant value: $\phi(t)= \phi_0$. This solution describes the field that drives a topological phase transition from a 4D Euclidean space to a 4D hyperbolic space-time.

\subsection{Quantum back-reaction}

In order to describe the exact back-reaction effects, we shall consider Relativistic Quantum Geometry (RQG), introduced in
\cite{rb}. In this formalism the manifold is defined with the connections\footnote{To simplify the notation we denote $\sigma_{\alpha} \equiv \sigma_{,\alpha}$.}
\begin{equation}\label{ga}
\Gamma^{\alpha}_{\beta\gamma} = \left\{ \begin{array}{cc}  \alpha \, \\ \beta \, \gamma  \end{array} \right\}+ \sigma^{\alpha} \hat{g}_{\beta\gamma} ,
\end{equation}
where $\delta{\Gamma^{\alpha}_{\beta\gamma}}=\sigma^{\alpha} \hat{g}_{\beta\gamma} $ describes the displacement of the Weylian manifold\cite{weyl} with respect to the Riemannian background, which is described by the Levi-Civita symbols in (\ref{ga}). In our approach, $\sigma(x^{\alpha})$ is a scalar field and the covariant derivative of the metric tensor in the Riemannian background manifold is null (we denote with a semicolon the Riemannian-covariant derivative): $\Delta g_{\alpha\beta}=g_{\alpha\beta;\gamma} \,dx^{\gamma}=0$. The variation of the metric tensor in the sense of (\ref{ga})\footnote{In what follows we shall denote with a $\Delta$ variations on the Riemann manifold, and with a $\delta$ variations on a Weylian-like manifold.}: $\delta g_{\alpha\beta}$, will be
\begin{equation}\label{gab}
\delta g_{\alpha\beta} = g_{\alpha\beta|\gamma} \,dx^{\gamma} = -\left[\sigma_{\beta} g_{\alpha\gamma} +\sigma_{\alpha} g_{\beta\gamma}
\right]\,dx^{\gamma},
\end{equation}
where
\footnote{We can define the operator
\begin{displaymath}
\check{x}^{\alpha}(t,\vec{x}) = \frac{1}{(2\pi)^{3/2}} \int d^3 k \, \check{e}^{\alpha} \left[ b_k \, \check{x}_k(t,\vec{x}) + b^{\dagger}_k \, \check{x}^*_k(t,\vec{x})\right],
\end{displaymath}
such that $b^{\dagger}_k$ and $b_k$ are the creation and destruction operators of space-time, such that $\left< B \left| \left[b_k,b^{\dagger}_{k'}\right]\right| B  \right> = \delta^{(3)}(\vec{k}-\vec{k'})$ and $\check{e}^{\alpha}=\epsilon^{\alpha}_{\,\,\,\,\beta\gamma\delta} \check{e}^{\beta} \check{e}^{\gamma}\check{e}^{\delta}$. }
\begin{equation}
dx^{\alpha} \left. | B \right> =  \hat{U}^{\alpha} dS \left. | B \right>= \delta\check{x}^{\alpha} (x^{\beta}) \left. | B \right> ,
\end{equation}
is the eigenvalue that results when we apply the operator $ \delta\check{x}^{\alpha} (x^{\beta}) $ on a background quantum state $ \left. | B \right> $, defined on the Riemannian manifold\footnote{In our case the background quantum state can be represented in a ordinary Fock space in contrast with LQG, where operators is qualitatively different
from the standard quantization of gauge fields.}. We shall denote with a {\it hat} the quantities represented on the Riemannian background manifold. The Weylian-like line element is given by
\begin{equation}
dS^2 \, \delta_{BB'}=\left( \hat{U}_{\alpha} \hat{U}^{\alpha}\right) dS^2\, \delta_{BB'} = \left< B \left|  \delta\check{x}_{\alpha} \delta\check{x}^{\alpha}\right| B'  \right>.
\end{equation}
Hence, the differential Weylian-like line element $dS$ provides the displacement of the quantum trajectories with respect to the "classical" (Riemannian) ones. When we displace
with parallelism some vector $v^{\alpha}$ on the Weylian-like manifold, we obtain
\begin{equation}
\delta v^{\alpha} = \sigma^{\alpha} g_{\beta\gamma} v^{\beta} dx^{\gamma}, \qquad \rightarrow \qquad \frac{\delta v^{\alpha}}{\delta S} =
\sigma^{\alpha} v^{\beta}\,g_{\beta\gamma}\, \hat{U}^{\gamma},
\end{equation}
where we have taken into account that the variation of $v^{\alpha}$ on the Riemannian manifold, is zero: $\Delta v^{\alpha}=0$. Hence, the norm of the vector on the Weylian-like
manifold is not conserved: $\frac{\delta v^{\alpha}}{\delta S} \frac{\delta v_{\alpha}}{\delta S}=-\left(\sigma^{\alpha} \hat{U}_{\alpha}\right) \left(v_{\gamma} \hat{U}^{\gamma} \right) \left(\sigma^{\nu} v_{\nu}\right) \neq 0$. However, the Weylian covariant derivative\cite{weyl} on the manifold generated by (\ref{ga}) is nonzero: $ g_{\alpha\beta|\gamma} = \sigma_{\gamma}\,g_{\alpha\beta}$. From the action's point of view, the scalar field $\sigma(x^{\alpha})$ is a generic geometrical transformation that leads invariant the action\cite{rb}
\begin{equation}\label{aac}
{\cal I} = \int d^4 \hat{x}\, \sqrt{-\hat{g}}\, \left[\frac{\hat{R}}{2\kappa} + \hat{{\cal L}}\right] = \int d^4 \hat{x}\, \left[\sqrt{-\hat{g}} e^{-2\sigma}\right]\,
\left\{\left[\frac{\hat{R}}{2\kappa} + \hat{{\cal L}}\right]\,e^{2\sigma}\right\},
\end{equation}
Hence, Weylian quantities will be varied over these quantities in a semi-Riemannian manifold so that the dynamics of the system preserves the action: $\delta {\cal I} =0$, and we obtain
\begin{equation}
-\frac{\delta V}{V} = \frac{\delta \left[\frac{\hat{R}}{2\kappa} + \hat{{\cal L}}\right]}{\left[\frac{\hat{R}}{2\kappa} + \hat{{\cal L}}\right]}
= 2 \,\delta\sigma,
\end{equation}
where $\delta\sigma = \sigma_{\mu} dx^{\mu}$ is an exact differential and $\hat{V}=\sqrt{-\hat{ g}}$ is the volume of the Riemannian manifold. Of course, all the variations are in the Weylian geometrical representation, and assure us gauge invariance because $\delta {\cal I} =0$.
The metric that takes into account the quantum back-reaction effects is
\begin{equation}\label{met1}
g_{\mu\nu} =  {\rm diag}\left[ \left(\frac{\pi a_0}{2}\right)^2 \frac{e^{2\sigma}}{\hat{\theta}^2} , - a^2(t) \,\left(\frac{\pi a_0}{2}\right)^2 \frac{e^{-2\sigma}}{\hat{\theta}^2}, - a^2(t) \,\left(\frac{\pi a_0}{2}\right)^2 \frac{e^{-2\sigma}}{\hat{\theta}^2}, - a^2(t) \,\left(\frac{\pi a_0}{2}\right)^2 \frac{e^{-2\sigma}}{\hat{\theta}^2}\right],
\end{equation}
with a volume: $V=\hat{V} e^{-2\sigma}$. The scalar curvature is altered due to the quantum back-reaction effects:
\begin{equation}
{R} - \hat{R} = - 3\left[ \nabla_{\mu}\sigma^{\mu} + \sigma_{\mu} \sigma^{\mu} \right],
\end{equation}
such that the $\nabla$-operator acts on the Riemman manifold.

\subsection{Energy density fluctuations}

The amplitude of energy density fluctuations are\cite{mb}: $\frac{1}{\hat{\rho}} \frac{\delta \hat{\rho}}{\delta S} = - 2 \left(\frac{\pi}{2a_0}\right) \, \hat\theta \sigma'$,
where for ${\sigma}' = \left< (\sigma')^2 \right>^{1/2}$, such that $\left< (\sigma')^2 \right> = \frac{1}{(2\pi)^{3}} \, \int^{\infty}_{2\sqrt{2}\epsilon/\pi} d^3k ({\xi}_k)' \, ({\xi}^*_k)'$.
Here, the modes $\xi_k$ must be restricted by the normalization condition: $({\xi}_k^*)' \xi_k - ({\xi}_k)' \xi^*_k= i \hat{\theta}^2 \left(\frac{2}{\pi a_0}\right)^2$, in order for the field $\sigma$ to be quantized\cite{rb}
\begin{equation}\label{con}
\left[\sigma(x), \sigma_{\mu}(y)\right] =i \, \hbar \Theta_{\mu} \delta^{(4)}(x-y).
\end{equation}
Here, $\Theta_{\mu}=\left[\hat{\theta}^2 \left(\frac{2}{\pi a_0}\right)^2,0,0,0\right]$ are the components of the background relativistic 4-vector on the Riemann manifold. The equation of motion for the modes of $\sigma$: $\xi_k(\hat\theta)$, is
\begin{equation}\label{mm}
\xi_k'' -   \frac{2}{\hat\theta} \xi'_k + k^2\, \xi_k(\hat\theta) =0,
\end{equation}
where the {\em prime} denotes the derivative with respect to $\hat\theta$.
The quantized solution of (\ref{mm}) results to be
\begin{equation}
\xi_k(\theta)= \frac{i}{2}\left(\frac{\pi}{2 a_0}\right) k^{-3/2} \,e^{-i k \hat\theta} \left[k\hat\theta-i\right].
\end{equation}
Therefore, since $\epsilon \ll 1$ and $k_0(\hat\theta)=\frac{\sqrt{2}}{\hat\theta}$, the amplitude of density energy fluctuations on super Hubble scales, become
\begin{equation}
\left|\frac{1}{\hat{\rho}} \frac{\delta \hat{\rho}}{\delta S}\right| =  \frac{\pi \epsilon^2}{4\sqrt{2} a^2_0} ,
\end{equation}
which is a constant.

\section{Conclusions}

We have solved in a natural manner the quantum to classical transition of the geometrical relativistic quantum fluctuations: $\sigma$, which has a geometric origin. The modes of $\sigma$ with wave number $k>k_0(\hat\theta)$ are stable and oscillate, but modes with $k<k_0(\hat\theta)$ are unstable. Notice that at the beginning of pre-inflation
$k_0(\hat\theta=\pi/2)={2\sqrt{2}\over \pi}$, so that almost all the spectrum is stable, but with the transition from an Euclidean to an hyperbolic space-time, the modes become unstable on almost all the range of the spectrum as $\hat\theta \rightarrow 0$. This quantum-to-classical transition can be seen in the evolution of the commutator (\ref{con}), which is proportional to $\hat\theta^2$, so that it becomes null with the expansion of the universe. This means that our formalism of pre-inflation describes in a natural manner the quantum-to-classical transition of the geometric relativistic quantum fluctuations $\sigma$. Therefore, the causal time emerges simultaneously with the classicality of the fluctuations. It can be seen in the Fig. (\ref{time}), where the $\Re{\left[\tau(t)\right]}$ is begins to increase with the expansion of the universe, but $\Im{\left[\tau(t)\right]}$ remains frozen. It is very interesting to notice that $\lim{\Re{\left[\tau(t)\right]}}_{t\rightarrow\infty}\rightarrow t$, so that, both $\Re{\left[\tau(t)\right]}$ and $t$ become indistinguishable with the increasing of the universe.\\

\section*{Acknowledgements}

\noindent  M. Bellini acknowledges UNMdP and  CONICET (Argentina) for financial support.\\

\end{document}